\documentclass[onecollarge,natbib]{svjour2}
\bibpunct{[}{]}{;}{n}{}{,} 
\smartqed  
\usepackage{graphicx}
\usepackage{epsfig}
\usepackage{graphicx,amsmath}
\newcommand\ba{\begin{eqnarray}}
\newcommand\ea{\end{eqnarray}}

\newcommand{\be}{\begin{equation}}
\newcommand{\ee}{\end{equation}}

\newcommand{\bas}{\begin{eqnarray*}}
\newcommand{\eas}{\end{eqnarray*}}

\def\sla#1{\rlap\slash #1}

\newcommand{\bno}{\begin{eqnarray*}}
\newcommand{\eno}{\end{eqnarray*}}

\journalname{Few-Body Systems}
\begin{document}
\title{In-Medium $K^+$ Electromagnetic Form Factor with a Symmetric Vertex in a Light Front Approach
\thanks{
Presented by George H. S. Yabusaki at LIGHT-CONE 2017 Conference in Mumbai, India.}
 }
\author{George H. S. Yabusaki$^{1,2}$ \and
        J. P. B. C. de Melo$^2$ \and
        Wayne de Paula$^1$ \and
        K. Tsushima$^2$ \and
        T. Frederico$^1$ }
\institute{$^1$Instituto Tecnol\'ogico de Aeron\'autica - ITA \at
              Pra\c ca Marechal Eduardo Gomes, 50, Vila das Ac\'acias, 12228-900, S\~ao José dos Campos, SP - Brazil \\
              \email{yabusaki@gmail.com}           
              \and
             $^2$Laborat\'orio de F\'isica Te\'orica e Computacional - UCS \at
              Rua Galv\~ao Bueno, 868, Bloco B - Sala 107, Liberdade, 01506-000, S\~ao Paulo, SP - Brazil \\
}
\date{Received: date / Accepted: date}

\maketitle

\begin{abstract}

Using the light-front $K^ +$-meson wave function based on a Bethe-Salpeter amplitude model for the quark-antiquark
bound state, we study the Electromagnetic Form Factor (EMFF) of the $K^ +$-meson in nuclear
medium within the framework of light-front field theory.
The $K^ +$-meson model we adopt is well constrained by previous and recent studies to explain its properties in vacuum.
The in-medium $K^ +$-meson EMFF is evaluated for the plus-component of the electromagnetic current, $J^+$, in the
Breit frame.
In order to consistently incorporate the constituent up and antistrange quarks of the $K^ +$-meson immersed in symmetric
nuclear matter, we use the quark-meson coupling (QMC) model, which has been widely applied to various hadronic
and nuclear phenomena in a nuclear medium with success.
We predict the in-medium modification of the $K^ +$-meson EMFF in symmetric nuclear matter.
It is found that, after a fine tuning of the regulator mass, i.e. $m_R = 0.600$ GeV, the model is suitable to fit the
available experimental data in vacuum within the theoretical uncertainties, and based on this we predict the 
in-medium modification of the $K^ +$-meson EMFF.


\end{abstract}

\section{Introduction}
\label{intro}
The main purpose of this work is to investigate
the in-medium modification of the $K^ +$-meson EMFF in symmetric
nuclear matter combined with the QMC model \cite{Nucleon}, where the $K^ +$-meson model \cite{Yabusaki} is adjusted
so as to provide the best description of the $K^ +$-meson EMFF data in vacuum.
The study of the lighter pseudoscalar mesons plays an important role to understand the low energy QCD. Their static and dynamical properties have also been investigated theoretically and experimentally \cite{Frederico92,Maris97,Choi98,Melo99,Choi,Fei,Krutov,Dally,Amendolia,PDG}.
With respect to the description of bound states on the light cone, a detailed review of hadronic wave functions in QCD-based models can be found in Ref.~\cite{Brodsky98}.
Additional important knowledge about the meson's internal structure can be inferred from their
valence-quark distribution functions. The theoretical framework we adopt is the light-front
field theory \cite{Brodsky98,Dirac}, more specifically, we use a symmetric vertex model for $\vert u\bar{s}\rangle$ $K^ +$-meson bound-state in the light-front approach for the Bethe-Salpeter amplitude. 
The light-front component $J^+$ of the electromagnetic current has been successfully 
used to calculate elastic form factor. 
For the symmetric $K-q\bar{q}$ vertex model~\cite{Pacheco2002}, the components of the current are 
conveniently obtained in the Drell-Yan
frame, where the light-front bound state wave functions are defined 
on the hypersurface $x^0+x^3 = 0$
and are covariant under kinematical boosts due to the stability of 
Fock-state decomposition~\cite{Brodsky98,Perry}.
In this work, we consider the symmetric vertex function with the intention
 to optimize and unify the parameter set to simultaneously reproduce 
the electromagnetic form factor. Our numerical results are compared with experimental 
data in vacuum up to $Q^2\approx 0.10~\text{GeV}^2$ to explore the validity of the model, where 
$Q^2=-q^2>0$, with $q$ being the four-momentum transfer.

\section{The Model}
\label{sec:1}
The electromagnetic current for a two-fermion bound state system with 
spin zero and intrinsic negative parity, $0^{-}$, a $K^+$-meson ($\vert u\bar{s}\rangle$ bound state), is calculated in 
one-loop approximation (triangle diagram shown in Fig. 1), modelling the Bethe-Salpeter 
amplitude through a symmetric vertex function 
in momentum space with a pseudoscalar coupling between $K^+$-meson and quarks.
This coupling is given by the effective Lagrangian \cite{Yabusaki,Pacheco2002,Pacheco2012}:
\begin{equation}
{\cal L}_{eff} = -\imath \frac{\hat{m}}{f_{K^+}}
\bar{q} \frac{1}{\sqrt{2}}(\lambda_4 + \imath  \lambda_5) \gamma^5 
q \frac{1}{\sqrt{2}}(\phi_4 -\imath \phi_5) \Lambda^*,
\end{equation}
here, $q=(u,d,s)^{T}$  and  $K^{+}=\frac{1}{\sqrt{2}}(\phi_4 -\imath \phi_5)$, $\hat{m}$ is given by the $\frac{m^{*}_{u} + m_{\bar{s}}}{2}$ with $m_{u}^{*}=m_{u}+V_s$ , $\Lambda^{*}$ the symmetric vertex function in nuclear medium and $f_{K^{+}}$ the $K^+$-meson decay constant. In this study we approximate to use $f_{K^{+}}$ value in vacuum.
In the Hartree mean field approximation the modifications enter as the shift of the light-quark momentum 
via $ P^{\mu} \rightarrow P^{*\mu}=P^{\mu}+V^{\mu} = P^{\mu }+\delta_{0}^{\mu}V^{0} $ due to the vector potential, and in the Lorentz-scalar part through the Lorentz-scalar potential $V_s$ as $m_{u} \rightarrow m_{u}^{*}=m_{u}+V_s$ \cite{Nucleon,Pion} and 
$m_{\bar{s}} \rightarrow m_{\bar{s}}^{* }=m_{\bar{s}}$ based on the QMC model \cite{Nucleon}. The QMC model has been applied to many nuclear and hadronic phenomena in a nuclear medium with success, and the inputs in vacuum as well as quantities calculated in-medium shown in Table 1 were adopted in the model to describe the effects of nuclear medium. The electromagnetic current for $K^+$-meson with the plus-component, is obtained from the covariant expression Eq.~(2) corresponding to the triangle diagram in Fig. 1
\vspace{-0.5cm}
\begin{figure*}[h]
\centering
\includegraphics[width=0.65\textwidth]{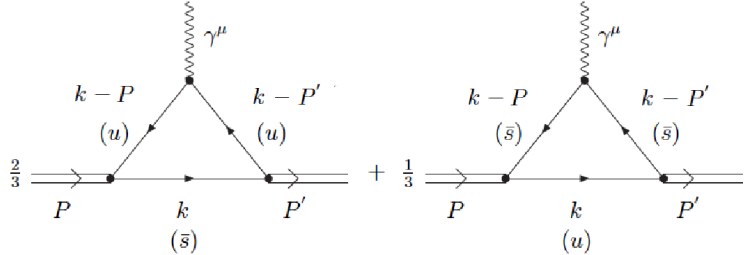}
\caption{Diagrams for $K^+$-meson photon interaction.}
\label{fig:diag01}
\end{figure*}
\vspace{-0.6cm}
\begin{eqnarray}
J^{\mu}(q^2) = -ie\frac{\hat{m}^{2}}{f^{2}_{K^+}}N_{C}
\int \frac{d^{4}k}{(2\pi)^4}\left\{\frac{2}{3}
Tr[S(k,m_{\bar{s}})\gamma^{5}S(k^{*}-P^{'},m_{u}^{*})
\gamma^{\mu}S(k^{*}-P,m_{u}^{*})\gamma^{5}] \right. + \nonumber \\
 \left. \frac{1}{3}Tr[S(k^{*},m_{u}^{*})\gamma^{5}S(k-P^{'},m_{\bar{s}})\gamma^{\mu}
S(k-P,m_{\bar{s}})\gamma^{5}]\right\}
\Lambda^{*}(k+V, P)~\Lambda^{*}(k+V, P^{'}).
\end{eqnarray}
where $(k^{*})^{\mu}=k^{\mu }+\delta_{0}^{\mu}V^{0}$, $\Lambda^{*}(k+V, P^{'})$ and $\Lambda^{*}(k+V, P)$ 
are the symmetric vertex function in nuclear medium, where the vertex function is given by \cite{Pacheco2002}
\begin{eqnarray}
\Lambda^{*}(k+V, P)=\frac{C}{(k+V)^{2}-m^{*2}_{R}+i\epsilon)}+\frac{C}{((P-k-V)^{2}-m^{*2}_{R}+i\epsilon)},
\end{eqnarray}
with $N_{C} = 3$ being the number of colors in QCD, and
\small{
\begin{eqnarray}
S(k^{*}-P,m^{*}_{u})=\dfrac{1}{(\sla{k^{*}}-\sla{\hspace{-.5ex}P}-m^{*}_{u}+i\epsilon)}
~\text{and}~S(k-P,m_{\bar{s}})=\dfrac{1}{(\sla{k}-\sla{\hspace{-.5ex}P}-m_{\bar{s}}+i\epsilon)},
\end{eqnarray}}
are corresponding to the up and antistrange quark propagators, respectively, in symmetric nuclear matter. 
In addiction we use $m^{*}_{R}=m_{R}$, the vacuum value.

We summarize here the light-front model in vacuum for the symmetric vertex function $\Lambda$ ($V=0$ and $m^{*}_{R}=m_{R}$) 
for the pseudoscalar bound states. 
Also, we work in the Breit frame and using light-front variables, $k^+=k^0+k^3$, $k^-=k^0-k^3$ and $k^\perp=(k^1,k^2)$, 
and one has
\begin{eqnarray}
q^{+} =-q^{-}=\sqrt{-q^2} \sin{\alpha},~q_{x}=\sqrt{-q^{2}}\cos\alpha,~q_{y}=0~\text{and}~q^2=q^{+}q^{-}-(\vec{q}_{\perp})^2,
\end{eqnarray}
where the Drell-Yan condition $q^+=0$ is recovered with $\alpha =0^o$ \cite{Melo99,Pacheco2002,Bakker}.
As is well known the $K^{+}$-meson form factor can be extracted from the covariant expression below:
\begin{eqnarray}
F_{K^{+}}(q^{2})=\frac{1}{e(P+P')^{\mu}} <P^{\prime}|J^{\mu}|P>.
\end{eqnarray}

If covariance and current conservation are fulfilled, one can obviously
use any frame and any nonvanishing component of the current to calculate the electromagnetic
form factor. 
In the light-front approach, however, besides the valence component of the electromagnetic current,
we can have the nonvalence contribuition or zero modes; thus in the light-front,
this two contributions enter in the full electromagnetic form factor: 
\vspace{-0.2cm}
\begin{eqnarray}
F_{K^{+}}(q^{2})=F^{(I)}_{K^{+}}(q^{2}, \alpha)+F^{(II)}_{K^{+}}(q^{2}, \alpha),
\end{eqnarray}
where $\alpha =0^{o}$, $F^{(I)}_{K^{+}}(q^{2}, \alpha)$ has the loop integration 
on $k^-$ constrained by $0\leq k^{+}< P^{+}$ (see the
light-front time-ordered diagram in Fig. 2 (a)) the valence $(u\bar{s})$, 
and $F^{(II)}_{K^{+}}(q^{2}, \alpha)$ has 
the loop integration on $k^-$ in the interval $P^{+}\leq k^{+}\leq P^{\prime+}$ 
(see Fig. 2 (b)) the nonvalence $\vert u\bar{s}\rangle$, pair production contributions 
with $q^+>0$, we use only the valence component, since the non-valence component goes to zero in the adopted framework (further see Ref. \cite{Yabusaki} for more details).

\vspace{-0.7cm}
\begin{figure*}[h]
\centering
\includegraphics[width=0.35\textwidth]{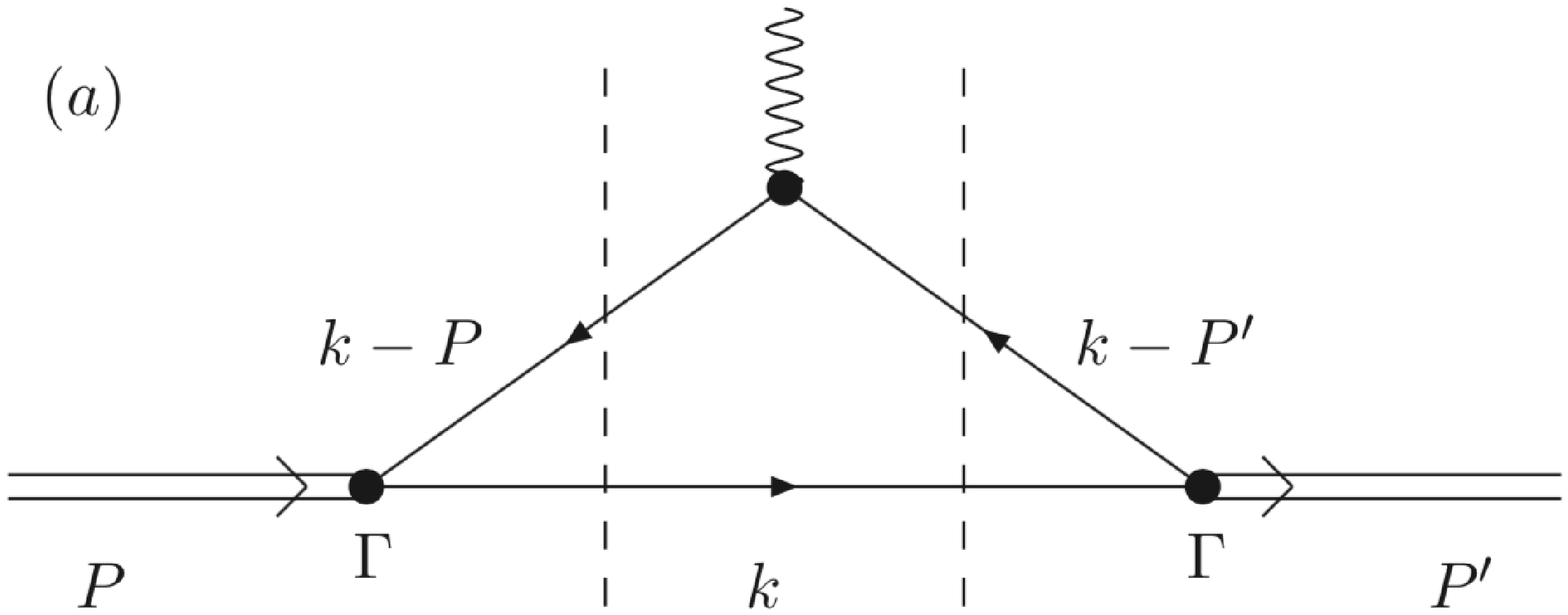}
\vspace{-0.5cm}
\label{fig:diag02}
\includegraphics[width=0.40\textwidth]{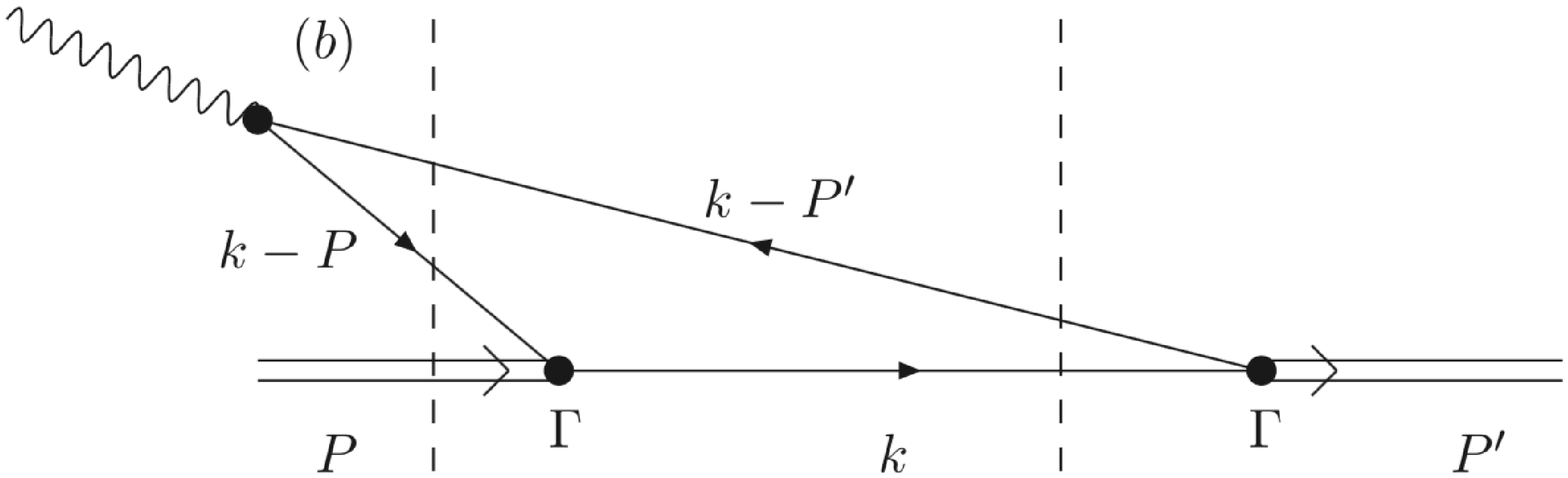}
\vspace{0.5cm}
\caption{Light-Front diagrams: (a) valence contribution and (b) non-valence contributions.}
\label{fig:diag03}
\end{figure*}
\vspace{-0.5cm}
Now we considerer in symmetric nuclear matter. After the $k^{*-}$ integration for $J^+$ current Eq.~(2) using  the Cauchy's Theorem, a light-front wave function emerges for a symmetric $\vert u\bar{s}\rangle$ vertex function with the change of variable $x^{*}P^{*+}=x{P^+}+V^{+}$ in medium where $x=\frac{k^+}{P^+}$. The $K^ +$-meson light-front wave function in symmetric nuclear matter is defined as:

\begin{eqnarray}
& \Phi^{*}(x^{*}, \vec{k}_{\perp})=
\frac{P^{+}}{m^{*2}_{K^{+}}-\mathcal{M}^{2}_{0}}
\left[\frac{\mathcal{N}}{(1-x^{*})(m^{*2}_{K^{+}}-\mathcal{M}^{2}_{0})}+
\frac{\mathcal{N}}{x^{*}(m^{*2}_{K^{+}}-\mathcal{M}^{2}_{R}))}\right] + [u\leftrightarrow\bar{s}],
\end{eqnarray} where $\mathcal{N}$ is the normalization factor, $\mathcal{M}^{2}_{0}$ is a mass operator
and $\mathcal{M}^{2}_{R}$ is a regulator mass function given by
{\small
\begin{eqnarray} 
\mathcal{M}^2_0=
 \frac{k_\perp^2+m_u^{*2}}{x^*}+\frac{(P-k)_\perp^2+m_{\bar{s}}^2}{(1-x^*)}-p_\perp^2~ \text{and}~
\mathcal{M}^2_R=
 \frac{k_\perp^2+m_u^{*2}}{x^*}+\frac{(P-k)_\perp^2+m_R^2}{(1-x^*)}-p_\perp^2,
\end{eqnarray}} and $[u\leftrightarrow\bar{s}]$ (recall that $m^{*}_{\bar{s}}=m_{\bar{s}}$). \\

Using only the valence component, the electromagnetic form factor evaluated in the Breit-frame reads \cite{Yabusaki,Melo99,Pacheco2002,Pacheco2012,Pion}
\begin{eqnarray}
F^{*(WF)}_{K^+}(q^2) = \frac{1}{2\pi^3(P^{*\prime+}+P^{*+})} 
\int \frac{d^{2}dk_\perp \theta(k^{*+})\theta(P^{*+}-k^{*+})}{k^{*+}(P^{*+}-k^{*+})(P^{*\prime+}-k^{*+})}
\Phi^{* \dagger}(x^{*}, \vec{k}_{\perp})\text{Tr}[\mathcal{O}^+]\Phi^{*}(x^{*}, \vec{k}_{\perp})+ [u\leftrightarrow\bar{s}],
\end{eqnarray}

the trace $\text{Tr}[\mathcal{O}^+]$ in light-front coordinates is

\begin{eqnarray}
 \text{Tr}[\mathcal{O}^+]&=&
  \tfrac{1}{4}k^{*+}q_\perp^2-\left(\frac{k^2_\perp+m_{u}^{*2}}{k^{*+}}\right) \, P^{*+}P^{*\prime+}-(P^{*\prime +}
      k_\perp\cdot P_\perp +P^{*+}k_\perp\cdot P_\perp^\prime) + [u\leftrightarrow\bar{s}].
 \end{eqnarray}

\section{Numerical Results}
We have three model parameters: the regulator mass, $m_R$, the quark masses, $m_u$ and $m_{\bar{s}}$ to compute EMFF. 
Our main aim of this work is to jointly analyze
the $K^{+}$-meson's elastic form factors to determine more accurately
the model's quark masses in view of future applications and to test whether a single mass scale, $m_R$, 
can satisfactorily describe experimental data for $K^{+}$-meson  \cite{Dally,Amendolia}, as well as to study the in-medium $K^+$ EMFF.
We use the quantities described in Tab. 1 for vacuum, and in-medium inputs computed by the QMC model \cite{Nucleon,Pion}. Our main results of this study are shown in Fig. 3.
As the nuclear matter density increases $K^+$ EMFF decreases faster than that in vacuum, and thus this implies $K^+$ charge radius increases in-medium.\\
\begin{table}[]
\centering
\begin{tabular}{|c|c|c|c|c|}
\hline
$\rho/\rho_0$              & $0.00$ & $0.25$ & $0.50$ & $0.75$\\ \hline
$m_K [GeV]$              & $0.494$ &$0.472$ & $0.453$ & $0.435$\\ \hline
$m_u~[GeV]$             &$0.220$  & $0.180$ & $0.143$ &$0.110$\\ \hline
$V \  \  \ [GeV]$           & $0.000$ & $0.029$ & $0.058$  & $0.088$\\ \hline
$m_{\bar{s}}~[GeV]$ & \multicolumn{4}{c|}{0.440} \\ \hline
$m_R~[GeV]$   & \multicolumn{4}{c|}{0.600} \\ \hline
\end{tabular}
\caption{Parameters used to compute the electromagnetic form factor for $K^{+}$-meson in medium, which are calculated by the from QMC model, (for details see Ref. \cite{Nucleon,Pion}).}
\end{table}
\begin{figure*}[h]
\centering
\includegraphics[width=0.52\textwidth]{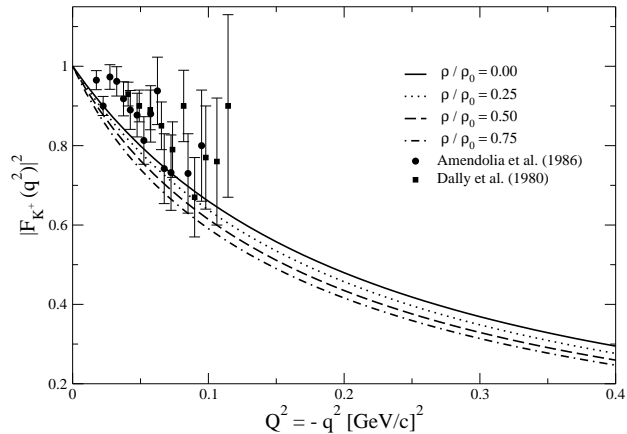}
\label{fig:f01}
\caption{Results for $K^{+}$-meson form factor squared calculated for different nuclear matter densities using the 
parameters from Tab. 1. Experimental data form vacuum
given by references \cite{Dally,Amendolia}.}
\end{figure*}

\footnotesize {\bf Acknowledgements:}
This work was partially supported by the Funda\c c\~ao de Amparo \`a Pesquisa do Estado de
S\~ao Paulo (FAPESP),~Brazil, No.~2015/16295-5 (JPBCM), and No.~2015/17234-0 (KT), 
~and Conselho Nacional de Desenvolvimento 
Cient\'ifico e Tecnol\'ogico~(CNPq), Brazil, No. 401322/2014-9 (JPBCM), 
No. 400826/2014-3 (KT), No. 308025/2015-6 (JPBCM), and No. 308088/2015-8 (KT).
This work was part of the projects, 
Instituto Nacional de Ci\^{e}ncia e Tecnologia - Nuclear Physics and Applications 
(INCT-FNA), Brazil, No. 464898/2014-5, and FAPESP Tem\'{a}tico, No. 2017/05660-0.

\end{document}